**Locality and Probability in Relativistic Quantum Theories and Hidden Variables Quantum Theories**

Avi Levy°, Meir Hemmo°

**ABSTRACT**

We use the framework of Empirical Models (EM) and Hidden Variables Models (HVM) to analyze the locality and stochasticity properties of relativistic quantum theories, such as Quantum Field Theory (QFT). First, we present the standard definition of properties such as determinism, No Signaling, Locality, and Contextuality for HVM and for EM and their relations. Then, we show that if no other conditions are added, there are only two types of EM: An EM is either classical, by which we mean that it is strongly deterministic, local, and non-contextual; Or else an EM is non-classical, in which case it is weakly deterministic, non-local and contextual. Consequently, we define a criteria for an HVM to be Lorentz invariant and prove that it implies No Signaling. As a result, we show that a Lorentz invariant and contextual (e.g. relativistic quantum theory) must be genuinely stochastic i.e., it cannot have a deterministic (strong or weak) HVM. This proof is an improved version of a theorem we proved previously, and it has a wider scope. Finally, we discuss Bell's definition of locality and show that it is equivalent to non-contextuality. We argue that Bell's justification for this definition tacitly assumes non-contextuality. We propose an alternative definition of locality for contextual and relativistic theories that accounts for correlations that result from common history and renders QFT a local theory.

I. **INTRODUCTION**

This paper is a continuation and extension of a previous paper by Levy and Hemmo (2020). In order for the present paper to be self-contained we repeat, in Section II 0in a concise and improved way the mathematical foundations and some of the results presented in the first paper. Then we proceed to extensions and new results.

° Philosophy Department, University of Haifa, email: levya@technion.ac.il.
° Philosophy Department, University of Haifa, email: meir@research.haifa.ac.il.

Empirical Models (EM) and Hidden Variables Models (HVM) are probabilistic frameworks for analyzing the properties of measurement scenarios that have been developed in a series of papers in the early 2000; see (Abramsky and Brandenburger 2011; Brandenburger and Yanofsky 2008; Brandenburger and Keisler 2016). In this framework certain properties of physical systems such as determinism (strong and weak), locality, no signaling and contextuality are defined in a formal mathematical way, and relations between them are derived. These mathematical definitions can be viewed as successors of Bell's (1964) pioneering approach of defining locality in a precise mathematical manner.

In this paper we focus on extending the EM and HVM framework to relativistic theories, by considering the measurement scenarios in Minkowski space-time. This enables us to derive conclusions about the stochastic nature and the locality property of QFT and of relativistic extensions of QM. The main contributions of this paper are:

- Introducing space-time coordinates to measurements in the EM and HVM framework and defining Lorentz invariance by requiring that all inertial frames have identical predictions and that temporal causality holds in every inertial frame.
  Then we provide a *novel* proof for the theorem, published in Levy and Hemmo (2020), according to which the Lorentz invariance property implies the No Signaling property (which in the case of a HVM is identical to Parameter Independence property; see below).
- Based on this result we prove a generalized version of a claim by Gisin (2011), which in the EM and HVM framework can be formulated as: an EM that is both contextual and Lorentz invariant cannot have an equivalent deterministic (weak or strong) HVM). Our new proof has the advantage that it overcomes an objection by Laudisa (2014) to proof by Gisin (2011). Our result implies that a contextual and Lorentz invariant EM must be *genuinely stochastic*.[1]
- From the analysis of the properties of EM we conclude that there are only two types of EM; one type is classical in the sense that it is non-contextual, local and strongly deterministic; the other is non-classical i.e., contextual, non-local and weakly deterministic.[2] It follows that the definitions of locality given in Bell (1964) and the definition of non-contextuality are equivalent. We analyze Bell's (1990) justification for his definition and show that he tacitly assumes non-contextuality.

---

[1] This conclusion generalizes the known fact that Bohm's (1952) hidden variables theory is committed to an absolute time order and therefore is not Lorentz invariant.
[2] These terms are defined in Section 0.

- Consequently we propose an alternative definition of locality that is based on No Signaling and on Lorentz invariant dynamics. Under this definition QM is indeed non-local (although it is No Signaling) but QFT is local as is expected from a relativistic theory.

The paper is organized in the following manner:

Section II presents the definitions of the EM and the HVM frameworks and in this context provides mathematical definitions for strong determinism, weak determinism, Parameter Independence, Outcome Independence and locality in the context of HVM models. Following these definitions, we list a set of logical relations holding between these properties and present a related existence theorem. Then, we discuss briefly the contextuality property and No Signaling property and other properties defined for EM.

In Section III, we review certain applications of the EM and HVM to QM theory that are useful in the subsequent sections: A condition for an EM to be non-contextual, a representation of the EPR scenario as an EM and an equivalent weakly deterministic HVM, a note about the relation between the Parameter Independence property and the No Signaling property and certain remarks about the interpretations of QM in the context of the EM and HVM framework.

In Section IV, we assign space-time coordinates for the measurements defined in the EM and HVM framework and we define an EM or a HVM to be Lorentz invariant if all inertial observers agree about the probabilistic predictions of the model and if temporal causality holds for every inertial observers. Then, we prove the main result of this paper stating that a Lorentz invariant EM is No Signaling and conclude that a contextual and Lorentz invariant EM cannot have an equivalent deterministic (week or strong) HVM; This no-go theorem asserts the known fact that Bohm's (1952) hidden variables theory cannot be extended to QFT without postulating a preferred inertial frame.

In Section V, we discuss Bell's definition of locality from Bell (1964) and argue, based on Sections II and III, that it is equivalent to the definition of non-contextuality. We show that Bell's (1990) justification of his locality definition implies that there exists a strongly deterministic HVM for the EPR scenario which is impossible for a contextual EM.. Therefore, we suggest that for contextual and relativistic theories this definition is too strict and propose instead an alternative definition of locality that is the conjunction of No Signaling and Lorentz invariant dynamic. We argue that this definition, which renders QFT (but not QM!) a local theory, is adequate for contextual and Lorentz invariant theories that are (as explained in Section IV) genuinely stochastic.

Section VI presents an interesting relationship between Bell's principle of local causality appearing in Bell (1990) and Reichenbach's principle of common cause. We explain how so-called "common cause systems" and the EM and HVM framework are related to each other.

Section VII sums up the conclusions of this paper.

Section VIII is an appendix presenting few examples of EM and HVM related to the discussion in the previous sections.

## II. THE EM AND HVM FRAMEWORK

In this section, we present in a concise way the framework of EM and HVM, presented more extensively in Sections 19.2-19.4 of (Levy and Hemmo 2020). Certain illustrative examples are given in Appendix A and an elaborated discussion can be found in Abramsky and Brandenburger (2011); Brandenburger and Yanofsky (2008); Brandenburger and Keisler (2016).

It is important to note that the EM and HVM framework does not pertain to *dynamical* models, hence certain definitions, especially those related to determinism and locality, are different from the definitions of these properties given in the context of dynamical theories. Also, these definitions constitute a formal mathematical framework and therefore applying their results to physical theories such as QM or QFT requires a careful philosophical analysis. For this reason, we present in this section the mathematical details only and delay the philosophical discussion of their implications for physical theories to Section III.

An Empirical Model (EM) consists of a set of admissible subsets of measurements with a corresponding set of outcomes, equipped with a consistent probability structure defined on these sets. In simple terms, an EM is a "table" of measurement subsets (that can be performed simultaneously) and their outcomes, equipped with the relative frequencies at which these outcomes are predicted to be observed.

Empirical Model (EM) (Abramsky and Brandenburger 2011, p. 5-7)

Let $X = \{A_1, A_2, \ldots, A_K\}$ [3] be a (finite) set of measurements and let $\{O_x\}_{x \in X}$ be the (finite) sets of their outcomes.

An <u>admissible</u> set of measurements subsets is a family $M$ of subsets of $X$ such that

1. It "covers" $X$: $\cup M = X$
2. It does not include subsets of its members: if $m_\alpha, m_\beta \in M$ and $m_\alpha \subseteq m_\beta$ then $m_\alpha = m_\beta$

An <u>outcome event</u> of a measurement subset $\{A_i, \ldots, A_k\} = m \in M$ is a specific element $\{o_i, \ldots, o_k\}$ of the space $O_i \times \ldots \times O_k$.

For each subset $m$ in $M$ there is a <u>consistent probability distribution</u> on the space $\prod_{i \in m} O_i$. We denote this condition by: $P_m \equiv P_{A_i, \ldots, A_k}$ where $m = \{i, \ldots, k\}$.

These probability distributions are <u>consistent</u> in the sense that for every subset of measurements in $X$ that is partial to two members $m_\alpha, m_\beta \in M$ the corresponding marginal probabilities are identical. Formally, if $\{A_i, \ldots, A_k\} \subset m_\alpha, m_\beta$ then $P^\alpha_{A_i, \ldots, A_k} = P^\beta_{A_i, \ldots, A_k}$, where $P^\gamma_{A_i, \ldots, A_k}$ denotes the marginal probability of $P_{A_i, \ldots, A_k}$ computed from $P_{m_\gamma}$.

<u>Definition</u>: An <u>Empirical Model</u> is defined by the sets $X$, $\{O_x\}_{x \in X}$ and $M$ together with the consistent family of probability distributions $\{P_m\}_{m \in M}$.

The more common term Hidden Variables Model (HVM) refers to an extension of EM, where a "state" space[4] is appended to the measurements and outcomes spaces with conditional probabilities for obtaining each outcome, given a state and a probability distribution defined over the state space.

---

[3] When there is no place for confusion the members of $X$ are denoted by their indices.
[4] The states of the HVM state space are constants in time, they represent the state of the system just before the measurement. Therefore, the HVM framework does not refer to the dynamics of the measured system.

Hidden Variables Model (HVM) (Abramsky and Brandenburger 2011, p. 23)

A HVM is an EM equipped with a "state" space $\Lambda$, a probability distribution $P_\Lambda$ defined on $\Lambda$ and a family of conditional probability distributions $\{P_{m|\lambda}\}_{m \in M}$ [5]. Note that the distribution $P_\Lambda$ does not depend on $m \in M$, which guarantees the property of $\lambda$ - independence (see Brandenburger and Yanofsky 2008).

These distributions assign for each "event" of every admissible measurement subset the probability $P_{m|\lambda}(o_i, \ldots, o_k | \lambda)$ $\lambda \in \Lambda$ and define in an obvious way a common probability distribution $P_{m,\Lambda}$ for every member $m \in M$.

A HVM is *equivalent* to an EM if one can describe the probabilities of the measurements' outcomes by summing over all the corresponding conditional probabilities of the states in the state space.

A HVM equivalent to an EM (Abramsky and Brandenburger 2011, p. 23)

An EM and a HVM defined by identical sets $X$, $\{O_x\}_{x \in X}$, $M$ are equivalent if $P_m$ is the marginal probability distribution of $P_{m,\Lambda}$ i.e., $P_m(o_i, \ldots, o_k) = \sum_\lambda P_{m|\lambda}(o_i, \ldots, o_k | \lambda) P_\Lambda(\lambda)$.

To conclude, an EM can be seen as the phenomenological description of experimental results, while the HVM adds a deeper structure that provides an "explanation" for the probabilities of the observed measurements' outcomes conditioned on the system's state just before the experiment is performed.

The EM and HVM framework allows defining, in a fully accurate mathematical way, important properties of experimental systems. Here is a list of some of the properties of HVM that came up in the literature.

---

[5] It is assumed that $P_\Lambda$ does not depend on the specific subset $m \in M$, a property that is known as "$\lambda$ independence". This assumption has implications for the existence theorem in Section 0.

Definitions of properties of HVM

Let us denote the outcomes of the measurements $A_i$ by $\{a_i^j\}$; the range of the upper index $j$ will be implicit from the context.

- A HVM is *Strongly Deterministic* (SD) if the state of the system determines uniquely the outcome of every measurement, i.e., $\forall i \in X$, $\forall \lambda \in \Lambda$, $\exists j$ s.t. $P_{A_i|\lambda}(a_i^j | \lambda) = 1$.

- A HVM is *Weakly Deterministic* (WD) if the state of the system determines uniquely the outcomes of every admissible series of measurements[6], i.e.,
$$\forall \{A_i, \ldots, A_k\} = m \in M, \; \forall \lambda \in \Lambda, \; \exists \; (a_i^j, \ldots, a_k^l) \text{ s.t. } P_{m|\lambda}(a_i^j, \ldots, a_k^l | \lambda) = 1.$$

- A HVM is *Parameter independent* (PI) if for every state, the outcome of a measurement does not depend on other measurements performed simultaneously with it, i.e.,
$$\forall i \in X, \; \forall j, \; \forall m_\alpha, m_\beta \in M \text{ s.t. } A_i \in m_\alpha, m_\beta, \; \forall \lambda \in \Lambda : P_{m_\alpha|\lambda}(a_i^j | \lambda) = P_{m_\beta|\lambda}(a_i^j | \lambda).$$

- A HVM is *Outcome Independent* (OI) if for every state, the outcome of a measurement does not depend on the outcomes of other measurements performed simultaneously with it, i.e.,
$$\forall \{A_i, \ldots, A_k\} = m \in M, \; \forall (a_i^j, \ldots, a_k^l), \; \forall \lambda \in \Lambda : P_{m|\lambda}(a_i^j, \ldots, a_k^l | \lambda) = P_{m|\lambda}(a_i^j | \lambda) \cdots P_{m|\lambda}(a_k^l | \lambda)$$

- A HVM is *(Bell) Local* if it is Parameter Independent and Outcome Independent.

The following relations between properties of HVM are provable (where '$\rightarrow$' and '$\leftrightarrow$' denote implication and logical equivalence, respectively):

(1) Strong Determinism $\rightarrow$ Weak Determinism $\rightarrow$ Outcome Independence (Brandenburger and Keisler (2016) p. 7, 10)

(2) Strong Determinism $\rightarrow$ Parameter Independence, hence (Brandenburger and Keisler (2016) p. 7)

(3) Strong Determinism $\rightarrow$ (Bell) Locality (Brandenburger and Keisler (2016) p. 7-8)

(4) Strong determinism $\leftrightarrow$ Weak Determinism and Parameter Independence (Brandenburger and Keisler (2016) p. 10), hence

(5) Strong determinism $\leftrightarrow$ Weak Determinism and (Bell) Locality (implied by (4), (1))

---

[6] Note that in strong determinism a *unique* outcome is determined for every measurement separately. This implies that the outcomes of admissible series of measurements are also unique, but the reverse is not necessarily true; see (Levy and Hemmo 2020) for more details and for examples.

An existence theorem[7]

For every EM, there exists an equivalent Weakly Deterministic HVM (Brandenburger and Yanofsky 2008, p. 9).

Definitions of properties of EM

An EM is <u>*Non-Contextual*</u> if there exists an equivalent *Strongly Deterministic* HVM (Abramsky and Brandenburger 2011, p. 9), and an EM is <u>*Contextual*</u> if it is not *Non-Contextual*.
It can be shown (Abramsky and Brandenburger 2011, p. 9) that this property is equivalent to the existence of a common probability distribution $P_{A_1,A_2,\ldots,A_K}\left(a_1^j, a_2^n, \ldots, a_K^l\right)$ from which the marginal distributions $P_m\left(a_i^j, \ldots, a_k^l\right)$ for every $m \in M$ can be derived. It follows that a *contextual* EM is one for which there is no such common probability distribution $P_{A_1,A_2,\ldots,A_K}$, or equivalently if there is no equivalent strongly deterministic HVM.

An equivalence theorem (Abramsky and Brandenburger 2011, theorem 8.1)
For an EM the two following properties are equivalent:
1. It has a (Bell) local equivalent HVM
2. The EM is non-contextual (it has a strongly deterministic HVM)

An EM is <u>*No Signaling*</u> (NS) if the outcome of a measurement does not depend on other measurements performed simultaneously with it, i.e.,
$\forall i \in X, \ \forall j, \ \forall m_\alpha, m_\beta \in M$ s.t. $A_i \in m_\alpha, m_\beta$: $P_{m_\alpha}\left(a_i^j\right) = P_{m_\beta}\left(a_i^j\right)$.

The definition of EM given in this section implies that every EM has the No Signaling property since from the consistency property of the family $\{P_m\}_{m \in M}$ of probability distributions it follows that $P_{m_\alpha}\left(a_i^j\right) = P_{A_i}\left(a_i^j\right) = P_{m_\beta}\left(a_i^j\right)$.

---

[7] Another existence theorem (Brandenburger and Yanofsky 2008) states that every EM has a strongly deterministic HVM. However, this theorem relies on the dependence of the state space distribution on the admissible measurement subsets ($\lambda$ dependence) which is excluded in the EM and HVM model presented in Abramsky and Brandenburger (2011).

The other properties defined above for HVMs, namely *strong determinism*, *weak determinism*, *Outcome Independence*, *Parameter Independence* and *locality*, can be extended to the corresponding EMs in the following way: *An EM has a certain property just in case there exists an equivalent HVM with this property.*

It is easy to see that the *Parameter Independence* property defined in this way is equivalent to the *No Signaling* property of EM.

These definitions for EM properties are consistent with other definitions appearing in the literature and they imply relations between these properties that resemble the relations between HVM properties mentioned above; see (Levy and Hemmo 2020) for additional details.

Conclusion:

By the existence theorem a *Contextual* EM (always) has an equivalent *Weakly Deterministic* HVM, but not a *Strongly Deterministic* one.

By the equivalence theorem a *contextual* EM cannot have an equivalent (Bell) *local* HVM.

It follow that there are only two possible scenarios for an EM (see Levy and Hemmo 2020, Section 3).[8]

An EM is either

- Classical (=Local=Strongly Deterministic=Non-Contextual), i.e., it has a Strongly Deterministic consistent HVM.

Or else:

- It is Non-Classical (=Non-Local=Contextual), i.e., it has a Weakly Deterministic consistent HVM, but not a Strongly Deterministic one.

This analysis shows that the *locality* property of EM is equivalent to the *non-contextuality* property.

### III. APPLICATION OF EM AND HVM TO QUANTUM MECHANICS

The EM and HVM framework is designed to describe measurements scenarios in QM. The EM setting in which not every series of measurements is admissible reflects the fact that in QM non-commuting observables cannot be measured simultaneously. The distinctions between weak determinism and strong determinism and between contextuality and non-contextuality originated in the different nature of measurement scenarios in QM experiments and classical experiments. Therefore, there is a reach literature that analyses the properties defined in Section II in various QM measurements scenarios and in other even more contextual (hypothetical) scenarios.

---

[8] The EM properties have relations resembling those described above for the HVM properties. These relations are skipped here but can be found in (Levy and Hemmo 2020).

As explained in Section II the EM and HVM framework captures only the static aspects of measurements scenarios and does not pertain to the dynamical aspects of the relevant theory. However, this limitation becomes an advantage when investigating the properties defined in Section 0 simultaneously for QM, its interpretations and relativistic extensions, and for QFT. This is due to the fact that despite their distinct dynamics all these theories share the same EM and some share also certain HVMs representing measurements' scenarios, such as the EPR scenario.

In this paper, we focus on the relativistic extension of the EM and HVM framework.[9] Let us start by presenting now a few issues that will be useful for our discussion of this extension in Section IV and to the analysis of locality in Section V.

A condition for an EM to be non-contextual

In the EM and HVM framework there is a simple algebraic procedure for verifying that an EM describing a measurement scenario is non-contextual. Since there is a strongly deterministic equivalent HVM for a non-contextual EM, it follows that there is a common probability distribution $P_X$ for the entire set of measurements $X$, i.e., for the random vector $P_{\{A_1, A_2, \ldots, A_K\}}$; see (Fine 1982; Brandenburger and Yanofsky 2008). This probability distribution defines a set of $\#\{O_i \times \ldots \times O_k\}$ that satisfies the constraint $\sum_{j_1, \ldots, j_K} P_X \left( o_1^{j_1}, \ldots, o_K^{j_K} \right) = 1$

Now, the consistent probability distributions for the admissible series $P_m$, $m \in M$ become marginal probability distributions of the common distribution $P_X$ and hence can be expressed (with a slight abuse of notation) as $P_{m=\{A_l,...,A_k\}}\left(o_l^{j_l},...,o_k^{j_k}\right) = \sum_{i\notin m, j_i} P_{\{A_1,A_2,...,A_K\}}\left(o_1^{j_1},...,o_K^{j_K}\right)$, where the summation is over all the outcomes of those measurements that are not included in the series $m$.

---

[9] A more conclusive discussion on implications of the EM and HVM framework for the understanding of QM is available in the references given in Section II0.

The set of these conditions for all the admissible series of measurements in $M$ can be written as a set of linear equations in the set of $\#\{O_i \times ... \times O_k\}$ unknown values $P_X\left(o_1^{j_1},...,o_K^{j_K}\right)$.

If this set of equations has a solution that satisfies $\sum_{j_1,...,j_K} P_X\left(o_1^{j_1},...,o_K^{j_K}\right) = 1$ and $P_X\left(o_1^{j_1},...,o_K^{j_K}\right) \geq 0$ for all the unknown values, then the EM is non-contextual.

Using this procedure makes it easy to verify that many known QM measurements' scenarios are contextual. See example 2 in Appendix A VIII for a proof that a version of the EPR experiment is contextual, based on this procedure.

Presenting the EPR experiment in the EM and HVM framework

The EPR experiment used in Bell (1964) to present his definition of locality and his celebrated inequality can be described by the following EM.

The set of measurements is $X = \{A_i, B_j\}$ $i,j = 1,...,K$, where $A_i$ are spin measurements performed by Alice, $B_j$ are spin measurements performed by Bob and the indices $i, j$ indicates the orientation setting of the measurements.

The set of admissible series of measurements includes only the couples $M = \{A_i, B_j\}_{i,j=1,...,K}$, since only one orientation can be used for a measurement at a time.

The sets of outcomes $\{O_x\}_{x \in X}$ are identical, so $\forall x \in X, O_x = \{-1,1\}$.

The family of probability distributions $P_{i,j}$ on the admissible series set are given by

$P_{i,j}(-1,-1) = P_{i,j}(1,1) = 0$, $P_{i,j}(-1,1) = p_{i,j}$, $P_{i,j}(1,-1) = 1 - p_{i,j}$, with $p \neq 0,1$.

As mentioned above, example 2 in the appendix shows that this EM (for $K = 2$) is contextual. It also presents a weakly but not strongly deterministic HVM, which is equivalent to the EPR EM as indicated by the existence theorem presented in Section 0.

The relation between the Parameter Independence property of HVM and No Signaling.

From the definitions given above of the Parameter Independence (PI) property for HVM and the No Signaling (NS) property for EM, it is clear that they are identical except for the condition on the parameter $\lambda$ appearing in the first. In the EM case the NS property is satisfied by definition as explained above. In the HVM case, if PI is not satisfied then there is at least one state $\lambda$ for which the distribution of certain measurement outcomes depends on the context i.e., on the identity of the measurement series

$$\exists i \in X, \; \exists j, \; \exists m_\alpha, m_\beta \in M \text{ s.t. } A_i \in m_\alpha, m_\beta \text{ and } P_{m_\alpha|\lambda}\left(a_i^j\right) \neq P_{m_\beta|\lambda}\left(a_i^j\right)$$

This means, at least in principle, that if the system is prepared in this state $\lambda$, then information can be transmitted to an observer performing the measurement $A_i$ by another observer just by choosing which measurement (other than $A_i$) she decides to perform. Thus, in the EPR experiment given by example 2 in Appendix A, the states $\lambda$=0,1,2,7 allow signaling.

QM and its interpretations as EM and as HVM

As demonstrated by the EPR example above the probabilistic predictions of QM for every measurement scenario (as long as the number of measurements and their outcomes are finite) can be represented by an EM i.e., a table of the outcomes of all admissible measurements series and their respective probabilities.

Born's rule of standard QM, based on the quantum state, is a stochastic HVM where the state parameter $\lambda$ is the quantum state just before the measurement. Note that the EM and HVM framework does not pertain to the dynamical aspect of QM, i.e., the Schrödinger equation. The only information required for the HVM definition is the quantum state just before the measurements and the coefficients of its representation as a series of eigenvectors of the measured observable.

When represented as an EM, standard QM is contextual since there are many measurements scenarios, such as the EPR experiment, which are contextual. However, as a HVM, with the quantum state and Born's rule, it has the property of Parameter Independence, i.e., no state in standard QM allows signaling. Formal proofs can be found in (Eberhard 1978, Abramsky and Brandenburger 2011).

Bohm's (1952) theory can be seen as a weakly deterministic HVM for QM. In Bohm's theory the system's state is described by the wavefunction and an additional position parameter for each particle. The wavefunction and the particles' positions evolve in time according to deterministic equations. Therefore, the result of every *position* measurement in Bohm's theory is completely determined, for each particle individually, by the system's state. Hence, it is strongly deterministic in relation to position measurements. However, when other properties such as spin are measured, the measurement scenarios are *contextual* and hence for these scenarios Bohm's theory is weakly deterministic but not strongly deterministic. The reason being that the state of the system (particles' positions and wavefunction) determines with certainty the measurement outcomes for the set of all spin values but not the individual spin value of every particle.

The GRW theory (Ghirardi, Rimini and Weber 1986) and its various extensions can also be considered as a HVM for the EM description of QM measurement scenarios. In this theory, the dynamical evolution of the state is based on modifying the linear dynamics of QM in order to introduce a stochastic collapse mechanism. But the GRW theory as a HVM is identical to the HVM (described above) representing standard QM, which is based on the quantum state and Born's rule. This is due to the fact that in the GRW theory the additional stochastic collapse mechanism does not change (in a detectable way) the probabilities of outcomes of QM measurements conditioned on the quantum state just before the measurement.

As for the Many Worlds interpretation, its stochastic structure is controversial, and in any case it does not propose any modification for the quantum state and Born's rule. Hence, as a HVM (and EM), for which the dynamical aspects are irrelevant, it is also identical to the HVM for standard QM.

## IV. <u>THE EM AND HVM FRAMEWORK FOR RELATIVISTIC THEORIES</u>

In order to utilize EM and HVM in relativistic theories, such as QFT, space-time locations should be introduced into this framework. This can be done by appending space-time coordinates to each measurement. Having done that, the relativistic relations of time-like, light-like and space-like are defined for each pair of measurements according to the relation between their corresponding coordinates. The No Signaling property for an EM now indicates that no superluminal communication is possible by performing measurements that are space-like separated.

Now, even though QM is not a relativistic theory, it does not allow superluminal signaling, hence, as a HVM it has the Parameter Independence (No Signaling) property as explained in Section III. On the other hand, if considered as an EM, QM is contextual and hence has a weakly, but not strongly, deterministic equivalent HVM. Such weakly deterministic HVM cannot satisfy No Signaling, since if it did, it would be strongly deterministic (by relation (4) of Section II) implying that QM is non-contextual.

Therefore, there is a seemingly paradoxical situation, where QM (as an EM) is No Signaling but it has a weakly deterministic HVM (e.g., Bohm's theory) that does not have this property and *must* have states that allow superluminal signaling as explained in Section III. The way out of this "paradox" is to assume that in every such HVM the "signaling" states *cannot* be prepared in practice. This is indeed the case in Bohm's theory since the position parameters are uncontrolled by the dynamics and are constrained to invariably have a distribution equal to the square of the amplitude of the quantum mechanical wavefunction.

We proceed now to consider the *static* relativistic extension of the EM and HVM framework. The term 'static' indicates that the measurements are equipped with space-time coordinates, but the states $\lambda \in \Lambda$ do *not* depend on space-time coordinates. In relativistic theories the Lorentz transformation between inertial frames may reverse the temporal order of events and hence the temporal order in which the measurements of an admissible subset are performed must not influence the probability distributions of this subset. In addition, to retain consistency, and since the measurements of an admissible series in an inertial frame are not necessarily performed at the same time, it is assumed that there is no external intervention that changes the probability distributions of the model during the time interval in which the whole series takes place.

On the basis of these assumptions, we make the following definition:

Lorentz invariant HVM (EM)

We define a HVM (with space-time coordinates attached to each measurement) to be *Lorentz invariant*, in the static sense, in case the following three conditions are satisfied:

1. All inertial frames agree on the probabilistic predictions of the HVM for *every* measurement.
2. In each inertial frame, the identity and order of the measurements performed at or after a time $t_2$ cannot affect the probability distributions pertaining to measurements performed before or at a time $t_1$, if $t_2 \geq t_1$ (temporal causality)[10].
3. Conditions 1 and 2 are satisfied for *every* assignment of coordinates to the set of measurements and for every state.

Note that since an EM can be considered as a HVM with a singleton state-space, this definition applies to an EM also.

Theorem

A Lorentz invariant HVM (EM) satisfies Parameter Independence (No Signaling)

Proof

We prove that the probability of the outcome of every measurement conditioned on the state of the system does not depend on the measurement subset in which it is performed. The proof for an EM is similar, referring to probabilities which are not conditioned on the state.

Stated in the notation of Section 0 we only need to prove the: following Lemma:

$\forall i \in X, \ \forall j, \ \forall m_\alpha, m_\beta \in M$ s.t. $A_i \in m_\alpha, m_\beta, \ \forall \lambda \in \Lambda : P_{m_\alpha | \lambda}\left(a_i^j\right) = P_{m_\beta | \lambda}\left(a_i^j\right)$ .

For clarity and brevity, the Lemma is proved for a simple HVM defined by $X = \{A, B, B'\}$, $M = \{\{A, B\}, \{A, B'\}\}$ and it will be clear from the proof that it can be easily extended to the general case.

Let us denote the outcomes of $A$ by $a$ and by condition 3 we can assume that $A$ is space-like separated from $B, B'$ and that (for simplicity) the measurements of $B, B'$ are performed (exclusively) in the same space-time location (see Figure 1). It follows that there *exists* an inertial frame, denote it by AB, in which the measurement of $A$ precedes the measurement of $B$ or $B'$.

---

[10] For example, the outcomes probability distribution of the A measurement must not depend on whether it is performed in the subsets ABC or ACB or ADB if B,C,D are performed after A.

Therefore, in the AB frame, $P_{A,B}^{AB}(a|\lambda) = P_{A,B'}^{AB}(a|\lambda) = P_A^{AB}(a|\lambda)$, since, by the time the measurement of *A* is performed, the measurement of *B* or *B'* has *not yet* been performed, and hence by condition 2, in this frame, the future choice between measuring *B* or measuring *B'* cannot influence the outcome of the *A*-measurement.[11] Since condition 1 implies that the equality $P_{A,B}(a|\lambda) = P_{A,B'}(a|\lambda)$ must hold in *every* inertial frame, this result entails Parameter Independence (No Signaling) for this specific HVM (EM). ☐

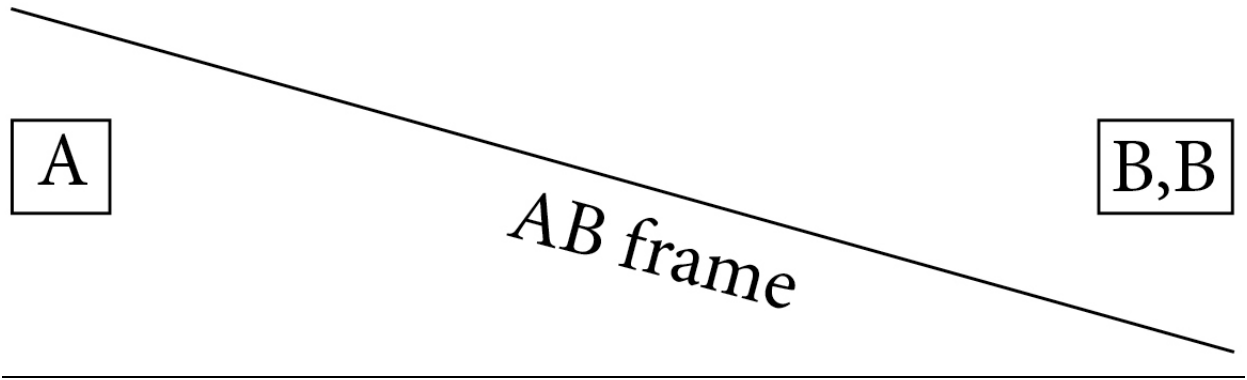

Figure 1: A Lorentz Invariant HVM must be No Signaling

Conclusion

A contextual EM has no equivalent Lorentz invariant HVM that is deterministic (weak or strong) and hence it is genuinely stochastic.[12]

Proof

Assume that the EM has an equivalent Lorentz invariant HVM that is weakly deterministic. By the theorem it is Parameter independence (No Signaling) and hence by relation (4) from Section

---

[11] Note that the outcome of the measurement of *A* can be correlated with the outcome of the measurement of *B* or *B'* via the state $\lambda$. However, the outcome of the *A*-measurement does not depend on whether *B* or *B'* have been performed, otherwise causality in the AB frame would be violated. Note that $\lambda$ independence is assumed and there is no probabilistic structure that selects (or "chooses") *B* or *B'*.

[12] This conclusion seems to contradict the existence theorem presented in Section 0 which guarantees the existence of a weakly deterministic HVM for every EM. What makes the difference is the additional Lorentz invariance requirement.

0 it is strongly deterministic. However, the existence of an equivalent strongly deterministic HVM makes this EM *non-contextual*, and this contradicts the assumption of the theorem. □

Comment

This conclusion generalizes a claim by Gisin (2011) which is presented in a narrower context. Also, the proof of the theorem avoids a comment by Laudisa (2014) to Gisin's proof, claiming that it is awkward. Laudisa's criticism is based on the use of two frames of references in Gisin's proof and claims that the non-local nature of the state parameter $\lambda$ is ambiguous. Also, he indicates that there is a difficulty in the proof when the two measurements are performed simultaneously. These two weaknesses are not relevant to our proof that considers only one frame of reference and uses the EM and HVM model where the parameter $\lambda$ is non-local by definition.

Note that our definition of a Lorentz invariant measurement scenario[13] is given explicitly in the context of the EM and HVM framework, which is static, i.e., it does not refer to the dynamical laws of the theory. QFT is Lorentz invariant in this static sense as well as in the more common dynamical sense which requires that the dynamical evolution over time of the system is Lorentz invariant. However, standard QM, although it is No Signaling, is not Lorentz invariant in this dynamical sense.

Since QFT is a quantum theory, it is contextual and as explained above, it is also Lorentz invariant. From the above Lemma and Conclusion we can derive our main result up to now: the physical reality described by QFT must be genuinely stochastic.[14] We stress that in this quite general form this is a novel result that generalizes other attempted proofs in the literature (e.g., by Gisin 2011; see above and Levy and Hemmo 2020).

As explained in Section III, Bohm's (1952) theory can be considered as a weakly deterministic HVM equivalent to QM. Since it is weakly deterministic, it cannot be Lorentz invariant in the sense defined above, and hence it cannot be compatible with QFT. This should not come as a surprise: it is well known that Bohm's theory is incompatible with special relativity, since its

---

[13] I.e., that all inertial frames agree on the probability distributions of the model.
[14] In other words, one may say in a somewhat metaphorical way, that God must play dice in a relativistic quantum world!

*dynamical* laws require absolute simultaneity and therefore the Bohmian dynamics can be self-consistent in a relativistic setting *only if* there is a preferred Lorentz frame of reference (see e.g., Albert 1992, Hemmo and Shenker 2013).[15]

Regarding the GRW theory, Esfeld and Gisin (2014) argued that neither GRW extensions (such as Tumulka's (2006) relativistic flash theory nor any other theory can account for the results of the Bell and EPR experiments in a Lorentz invariant manner. Their argument consists in fact of two different arguments.

The first one is based on considerations resembling those considered in (Gisin 2011). This argument *assumes* that there is a function that, given the system's state $\lambda$, determines the result of the measurements deterministically. However, as explained in Section III, the GRW (1986) theory is a stochastic HVM with probabilistic structure which is equivalent to that of standard QM. Hence, given the state space of the GRW theory *there is no such function*. Of course, there exists a weakly deterministic HVM which is empirically equivalent to the GRW theory (see Section 0) that has this property and is indeed not Lorentz invariant. But this HVM has a different state space.

So contrary to Esfeld and Gisin (2014), this static argument (referring to the states just before the measurement) which leads to their equations (1) and (2), does not entail that the GRW theory cannot account for the Bell (and EPR) quantum correlations in a Lorentz invariant way.

Esfeld and Gisin's (2014) second argument is based on *dynamical* considerations concerning the way in which the actual distribution of outcomes, satisfying the quantum correlations in Bell and EPR type experiments, *come about*. In this second argument, they claim that there must be a fact of the matter as to whether the occurrence of an outcome in one region of space-time depends only on that (local) region, or whether it also depends on the outcome occurred at space-like separation. But if there are such matters of fact about these dependencies, then (they argue) there must be some preferred Lorentz frame (and a preferred foliation of space-time), since these facts require an absolute time order. Therefore, if there are such facts, the Tumulka GRW flash theory cannot be genuinely relativistic. However, whether or not the GRW-Tumulka relativistic flash theory must acknowledge such facts is challenged by Tumulka (2009). We don't go into this debate here, because the question of whether or not such facts cannot be accounted for in a Lorentz invariant way addresses dynamical aspects of the GRW flash theory, and these aspects exceed the HVM and EM models, on which we focus here.

---

[15] Whether or not it is possible to discover empirically the preferred frame in Bohm's theory depends on some further issues concerning the meaning of *probability zero* given Bohm's statistical postulate; see (Hemmo and Shenker 2013).

The last three paragraphs above are meant to demonstrate what can be concluded from the relativistic extension of the EM and HVM framework and what cannot be claimed based on this *static* analysis. It does not refer to the vast area of research investigating the possibility of extending the dynamical aspects of theories like Bohm's (1952) and GRW (1986) to the relativistic domain. The interested reader is referred to (Jones, Guaita T., and Bassi 2021; Tumulka 2021) which address the GRW relativistic flash theory, and to Myrvold (2021) for a more general discussion.

## V. **BELL'S LOCALITY AND RELATIVISTIC EM AND HVM**

In this section we focus on the question of whether Bell's (1964, 1990) definition of locality, which is equivalent to the conjunction of Parameter Independence (No Signaling) and Outcome Independence, is adequate for contextual and relativistic theories.

In the language of the EM and HVM framework the locality property defined by Bell (1964, 1990) can be formulated as follows: an EM model is local if it has an equivalent local HVM. In this framework Bell showed that such a HVM cannot exist for the EPR scenario since it would violate Bell's inequality, which must be satisfied by a local HVM. Now, our main point in this section is the following: The equivalence theorem, presented in Section II, states that an EM that has an equivalent Bell's local HVM must be non-contextual and hence has an equivalent strongly deterministic HVM. It follows that requiring Bell's locality is equivalent to requiring non-contextuality for the EM[16] corresponding to the EPR setting, or in other words Bell's locality definition is equivalent to *strong determinism*!

---

[16] The concepts of contextuality and strong determinism as defined in the EM and HVM framework were presented only decades latter.

To see this, let us consider Bell's definition of locality for a HVM as the conjunction of Parameter Independence (PI/NS) and Outcome Independence (OI). After a first attempt to define local causality in (Bell 1990, Section 7), Bell proposes a refined definition in Section 9 of this paper (see Figure 2):

"… And let $\lambda$ denote any number of hypothetical additional complementary variables needed to complete quantum mechanics in the way envisaged by EPR Suppose that the c and $\lambda$ together give a complete specification of at least those parts of 3 blocking the two backward light cones."

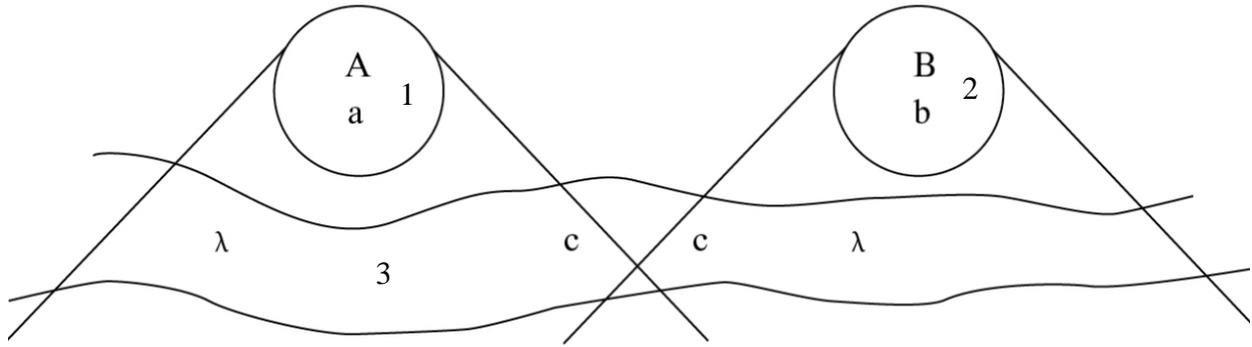

Figure 2: Spacetime setting and local causality for Bell-like measurements

"… By a standard rule, the joint probability can be expressed in terms of conditional probabilities: (9) $P_{A_\alpha B_\beta}(a,b|\lambda) = P_{A_\alpha B_\beta}(a|b,\lambda) P_{A_\alpha B_\beta}(b|\lambda)$"

" Invoking local causality, and the assumed completeness of $\lambda$ in the relevant parts of region 3, we declare redundant certain of the conditional variables in the last expression, because they are at space-like separation from the result in question. Then we have

(10) $P_{A_\alpha B_\beta}(a,b|\lambda) = P_{A_\alpha}(a|\lambda) P_{B_\beta}(b|\lambda)$"

As Bell explained (1990, Section 7) local causality here means that the outcome of the measurement of $A_\alpha$ cannot depend on the setting of the measurement $B_\beta$ (and vice versa), if these settings are not affected by causes in the intersection of the measurements' past light cones. However, they can be affected by the common parameter $\lambda$ (we absorb the parameter c into $\lambda$ for the sake of simplicity) which is not local and intersects both past light cones, as can be seen in Figure 6.

We note that for a HVM, local causality is equivalent to the Parameter Independence property, since it implies $P_{A_\alpha}(a|\lambda) = P_{A_\alpha B_\beta}(a|\lambda)$ for every value of the parameter $\beta$, and similarly for $P_{B_\beta}(b|\lambda) = P_{A_\alpha B_\beta}(b|\lambda)$. (In the definition of EM local causality is guaranteed by the compatibility assumptions on the admissible sets of measurements.)

Proposition:

If a HVM has both the local causality property and the completeness of $\lambda$, then (10) can be derived with certainty from (9) only for the case where there exists an equivalent HVM with a parameter $\tilde{\lambda}$ that defines the outcomes of both experiments in a strongly deterministic way.

Proof:

Let's look first at the following trivial counterexample were $\{a\} = \{b\} = \{+1, -1\}$, $\Lambda = \{0\}$ and

$$P_{A_\alpha B_\beta}(+1, -1 | 0) = \tfrac{1}{2}, P_{A_\alpha B_\beta}(-1, +1 | 0) = \tfrac{1}{2}$$

it follows that $P_{A_\alpha}(+1|0) = P_{A_\alpha}(-1|0) = \tfrac{1}{2}, P_{B_\beta}(+1|0) = P_{B_\beta}(-1|0) = \tfrac{1}{2}$.

but $P_{A_\alpha B_\beta}(a, b | 0) \neq P_{A_\alpha}(a|0) P_{B_\beta}(b|0)$.

In this example local causality is satisfied because one can add other experiments $A_{\alpha'}, B_{\beta'}$ with identical joint probability distributions, so that the parameters $\alpha, \beta, \alpha', \beta'$ have no impact. Also $\lambda$ defines completely the probability distributions of the experiments' outcomes, but not in a deterministic way. Despite the local causality and $\lambda$ completeness, equation (10) is not satisfied. To complete the example, one can see that if one takes $\tilde{\Lambda} = \{0, 1\}$ which defines deterministically the outcomes of the measurements by $P_{A_\alpha B_\beta}(+1, -1 | \tilde{\lambda} = 0) = 1, P_{A_\alpha B_\beta}(-1, +1 | \tilde{\lambda} = 1) = 1$,

then:

$$P_{A_\alpha B_\beta}(a, b | \tilde{\lambda}) = P_{A_\alpha}(a | \tilde{\lambda}) P_{B_\beta}(b | \tilde{\lambda})$$

and (10) is satisfied.

In the general case if a HVM satisfies (10) $P_{A_\alpha B_\beta}(a,b|\lambda) = P_{A_\alpha}(a|\lambda) P_{B_\beta}(b|\lambda)$, then it has, by definition, the Outcome Independence property and as explained above local causality is equivalent to Parameter Independence. Hence a HVM that satisfies local causality and $\lambda$ completeness and (10) satisfies both Parameter Independence Property and Outcome Independence. According to the results of Section II, there is an equivalent strongly deterministic HVM, and for this HVM equation (10) is always true. □

It follows that without the assumption that the HVM is strongly deterministic Bell's derivation of (10) from (9) based on local causality and $\lambda$ completeness is not always valid. Therefore, Bell's locality definition is equivalent to strong determinism. However, strong determinism cannot be satisfied by QM since it is contextual. A contextual EM can satisfy either Parameter Independence or Outcome Independence but not both. It follows that by definition every contextual theory cannot be Bell's local. It seems to us that a less strict definition of locality may be more useful for a contextual and relativistic theory.

A proposed definition of locality for contextual and relativistic theories

We propose that *relativistic locality* (RL) for a physical theory should be defined as the conjunction of the following two conditions:

1. The condition that the Parameter Independence (No-Signaling) property is satisfied by the static aspect of the theory, which is captured by the HVM / EM description; and
2. The condition that either the probabilistic structure of the theory does not depend on the dynamics of the theory or that it is preserved by *Lorentz invariant* dynamics.[17]

Note that the second condition refers to the dynamical aspect of the theory, which is not addressed by the EM and HVM framework. This condition is added to the definition of RL since in those cases where the probabilistic structure depends on the dynamics of the theory the dynamical RL condition is *necessary* for the static RL property to hold. The reason being that the dynamical RL condition guarantees that the probabilistic structure of the theory is preserved by the relativistic nature of the dynamical laws.

---

[17] Which is that the dynamical laws of the theory for inertial frames of reference satisfy the Lorentz transformations group.

The RL definition, which avoids Outcome Independence, has the advantage that it can account for correlations between measurement outcomes in both non-contextual (i.e., classical theories[18]) and contextual scenarios (i.e., non-classical theories, such as QM and QFT).

Comments:

1. The Parameter Independence (PI) / No Signaling (NS) property of a HVM / EM is strictly weaker than the static Lorentz invariance (LI) property defined in Section IV, where it is proved that (static) LI implies PI. The reverse implication fails, since (it is easy to show that) the probability distribution of a measurements scenario may be changed in such a way that the PI property is conserved but LI is lost. Our definition requires the PI / NS property, since the latter guarantees that temporal causality holds and in our view this is a condition that must be satisfied by a local theory.
2. The condition of Lorentz Invariant (LI) dynamics does not imply the static LI property defined in Section IV, since the static LI property requires the temporal causality condition which is not part of the LI dynamical condition.

According to the definition of RL, QFT is a *local* theory but QM is non-local, although it has the No Signaling property, since its dynamical laws are not Lorentz invariant.[19] Of course, both QM and QFT are non-local according to the definition of locality proposed by Bell.

As explained in Section IV, the physical reality described by QFT[20] is contextual and relativistic and hence genuinely stochastic. We suggest that in a genuinely stochastic theory such as QFT some sort of a non-local *"collapse" of the probabilistic structure must* occur, otherwise the theory is not genuinely stochastic. It seems to us that this "collapse" must be *primitive* in the sense that it cannot be further described by some underlying deterministic process (or by a deterministic HVM), since if it could, the theory would not be genuinely stochastic.[21]

---

[18] In classical scenarios, correlation effects are *masked* by non-contextuality, since such scenarios invariably have a strongly deterministic HVM, which is therefore No Signaling and outcome independent, and therefore (Bell's) local.
[19] The Schrödinger equation is not Lorentz invariant and may lead eventually to a disagreement between inertial frames about the probabilistic predictions of the theory. However, this aspect cannot be described and analyzed by the EM and HVM framework, in which the states are time-independent constants.
[20] Here we set aside the measurement problem in QFT.

## VI. REICHENBACH'S COMMON CAUSE PRINCIPLE AND THE EM AND HVM FRAMEWORK

There is an interesting relationship between the local causality requirement used by Bell to derive his definition of locality and the principle of common cause introduced by Reichenbach (1956). A framework for analyzing this relationship is presented in (Hofer-Szabó, Rédie and Szabó, 2013, p. 143-146) where a generalization of Reichenbach's principle named common cause system is defined in the context of an EPR experiment scenario.

We will now present some of the definitions and results from (Hofer-Szabó, Rédie and Szabó, 2013, Ch. 9) which are relevant to the present discussion and explain the correspondence between them and the EM and HVM framework.

A common cause scenario (for the EPR setting)

Let $(X, S, p)$ be a classical probability space and let $\{A_i, B_j\} \subseteq S$ $i \in I, j \in J$ denote the measurement events $\{A_i, B_j\}$ with corresponding outcomes $a_i, b_j \in \{0, 1\}$.

Assume that:

(1) $A_i \cap A_{i'} = \emptyset$, $B_j \cap B_{j'} = \emptyset$ for $i \neq i', j \neq j'$ and

(2) $p(a_i | A_i \cap B_{j'}) = p(b_i | A_i \cap B_j), p(a_i | A_{i'} \cap B_j) = p(b_i | A_i \cap B_j)$

(3) $\bigcup_i A_i = \bigcup_j B_j = X$

(4) $p(A_i \cap B_j) = p(A_i) p(B_j)$

Comments:

(i) This setting seems more general than the EM framework since the measurements $\{A_i, B_j\}$ are defined as events in a common classical probability space. However, the frameworks are practically equivalent since the probabilities $p(A_i)$, $p(B_j)$ of performing the measurements play no role in the results presented below and can be added formally to the EM framework with no impact on the results of Section II.

---

[21] Whether or not the resulting "collapse" theory will be compatible with special relativity is a question that goes beyond the scope of this paper.

(ii) The admissible subsets are defined by condition (1) above which implies that only the pairs $\{A_i, B_j\}$ can be measured.

(iii) Condition (2) guarantees that the consistency property of the EM is satisfied for all admissible subsets.

(iv) Condition (3) states that one admissible subset of measurements is performed with certainty, which is consistent with the EM framework.

Definition: a common cause system (definition 9.5 in Hofer-Szabó, Rédie and Szabó, 2013, p. 146).

Let $(X, S, p)$ be a classical probability space. A partition $\{C_k^{i,j}\}_{k \in K(i,j)}$ of S is a *common cause system* for the correlating pairs $\{A_i, B_j\}$ if for all $k \in K(i, j)$

$$p(a_i \cap b_j \mid C_k^{i,j}) = p(a_i \mid C_k^{i,j}) p(b_j \mid C_k^{i,j}).$$

As indicated in (Hofer-Szabó, Rédie and Szabó, 2013) the notion of a common cause system corresponds to the notion of "hidden variables" in the literature pertaining to the EPR correlations. Due to the additional probabilistic structure in the definition of this notion there is another requirement which is satisfied automatically in the EM framework which is the following.

No-conspiracy condition

A common cause system $\{C_k^{i,j}\}_{k \in K(i,j)}$ for the events $\{A_i, B_j\}$ satisfies the *no-conspiracy condition* if $p(A_{i'} \cap C_k^{i,j}) = p(A_{i'}) p(C_k^{i,j}), p(B_{j'} \cap C_k^{i,j}) = p(B_{j'}) p(C_k^{i,j})$ (see Hofer-Szabó, Rédie and Szabó, 2013, p. 147). A corresponding notion of $\lambda$ independence in the context of the EM and HVM can be found in (Brandenburger and Yanofsky 2008).

The main no-go result presented in (Hofer-Szabó, Rédie and Szabó, 2013, p. 149) is this: Define a common cause framework with $(X, S, p)$ a classical probability space, $CH = \{(1,3), (1,4), (2,3), (2,4)\}$ and $\{A_i, B_j\}_{(i,j) \in CH}$, and with probabilistic structure derived from the results of the EPR experiments. Then there is no extension of the probability space $(X, S, p)$ that contains a no-conspiracy common cause system for all pairs in $CH$.[22]

The correspondence of the common cause scenario for Bell's locality definition and the EM and HVM framework is now evident:

(i) A common cause scenario corresponds to an EM as explained above.

(ii) The extension of the space $(X, S, p)$ corresponds to constructing a HVM that is equivalent to the EM. The HVM is constructed by adding a state space $\Lambda$ to the EM together with a probability distribution $P_\Lambda$ and conditional probabilities $\{P_{m|\lambda}\}_{m \in M}$.
The no-conspiracy condition corresponds to the fact that the distribution of probabilities $P_\Lambda$ defined on $\Lambda$ does not depend on the admissible subset $m \in M$.

(iii) The partition $\{C_k^{i,j}\}_{k \in K(i,j)}$ of the extended space $(X', S', p')$ corresponds to the $\sigma$ algebra generated by the random variable defined by the state $\lambda$ in the HVM. Thus the conditional probabilities $p(a_i | C_k^{i,j}), p(b_j | C_k^{i,j})$ correspond to the conditional probabilities $p(a_i | \lambda), p(b_j | \lambda)$.

(iv) The definition of the common cause system stating that
$p(a_i \cap b_j | C_k^{i,j}) = p(a_i | C_k^{i,j}) p(b_j | C_k^{i,j})$ corresponds to requiring that the HVM satisfies both

---

[22] An extension of a probability space $(X, S, p)$ is a probability space $(X', S', p')$ with an injective Boolean algebra homeomorphism (see definition A.6 in (Hofer-Szabó, Rédie and Szabó, 2013, p. 18, in which: $h: S \to S'$ such that $\forall A \in S\ p(A) = p'(h(A))$.

Par*ameter Independence* and *Outcome Independence*, and therefore $p(a_i \cap b_j | \lambda) = p(a_i | \lambda) p(b_j | \lambda)$ which is Bell's (1964) definition of locality. Finaly:

(v) The no-go result stated above is a special case (pertaining to the EPR experiment) of the conclusion in Section IV according to which a contextual EM can't have a (Bell) local HVM.

We conclude that requiring the existence of a common cause system as defined in (Hofer-Szabó, Rédie and Szabó, 2013, Ch. 9, Definition 9.5) is equivalent to requiring that the common cause system is (Bell) local and hence non-contextual. Therefore, no contextual scenario (e.g. the EPR experiment) can have a common cause system.

## VII.    CONCLUSIONS

In a relativistic extension of the EM and HVM framework we showed that static Lorentz invariance[23] implies Parameter Independence (No Signaling). From this assertion we concluded that a contextual and Lorentz invariant theory (e.g., QFT) must be genuinely stochastic, i.e., it cannot have a deterministic, weak, or strong, compatible (i.e., empirically equivalent) HVM.

We further argued that Bell's definition of locality of a HVM is equivalent to requiring strong determinism and therefore it is equivalent to non-contextuality of the relevant EM, and is therefore too strict for a relativistic quantum theory. We proposed to define locality for a contextual and relativistic theory by the condition of Relativistic Locality (RL), which is the conjunction of No Signaling and Lorentz invariance of the dynamics. Under this condition, QFT (but not QM) is a local theory.

In future research, we will investigate the possibility of interpreting QFT, which as we argued above, is a genuinely stochastic theory, by a *combined* fields and events ontology. This ontology consists of a "hidden" stochastic layer, determined by fields, and a "detectable" irreversible events layer, in accordance with the ontology proposed by Haag (1996, Chapter VII). In this future work we will investigate the implications of the genuine stochastic nature of QFT with respect to the measurement problem (which we did not discuss here) and special relativity.

---

[23] In the sense we defined, namely that inertial observers must agree on the probability of every event and that in each frame of reference causal locality is satisfied.

## VIII. APPENDIX A

Example 1: A HVM equivalent to a classic EM which is No Signaling but outcome dependent

A HVM equivalent to a classical EM which is No Signaling but outcome dependent

Consider an EM and it's equivalent HVM with a singleton $\Lambda$ defined by:

$X = \{A, B\}, M = \{(A, B)\}, O_A = O_B = \{0,1\}$,

| $A, B$ | 0,0 | 0,1 | 1,0 | 1,1 |
|---|---|---|---|---|
| $P(A,B)=$ | 0 | ½ | ½ | 0 |

This HVM is not OI because $P(A,B) \neq P(A)P(B)$ but it is PI since there is only one admissible series. The EM is classical (non-contextual) since it is equivalent to the strongly deterministic HVM with two states $\Lambda = \{0,1\}$ each with $P(\lambda) = \frac{1}{2}$ and

| | $A, B$ | 0,0 | 0,1 | 1,0 | 1,1 |
|---|---|---|---|---|---|
| | $\lambda = 0$ | 0 | 1 | 0 | 0 |
| | $\lambda = 1$ | 0 | 0 | 0 | 1 |

This trivial example shows that when $A+B=0$ a HVM may not satisfy OI even when the equivalent EM is classical regardless of where the measurements are performed in space-time.

Example 2: The EPR experiment is contextual and has a weakly but not strongly deterministic equivalent HVM

The EM representing a version of the EPR scenario is defined in Section III by the following table:

| | 1,1 | 1,−1 | −1,1 | −1,−1 |
|---|---|---|---|---|
| $AB$ | 0 | ½ | ½ | 0 |
| $A'B$ | ⅛ | ⅜ | ⅜ | ⅛ |
| $AB'$ | ⅛ | ⅜ | ⅜ | ⅛ |
| $A'B'$ | ⅜ | ⅛ | ⅛ | ⅜ |

Assume that this EM is non-contextual, then according to the criteria presented in Section III there is a common probability distribution:

| $ABA'B'$ | 1,1,1,1 | 1,1,1,−1 | $\cdots$ | −1,−1,−1,−1 |
|---|---|---|---|---|
| | $p_1$ | $p_2$ | $\cdots$ | $p_{16}$ |

Let us look at the following 4 equations:

$P(AB = 1\,{-}1):\quad p_5 + p_6 + p_7 + p_8 = \tfrac{1}{2}$

$P(AB' = -1\,{-}1):\ p_7 + p_8 + p_{15} + p_{16} = \tfrac{1}{8}$

$P(A'B = 1\,1):\ p_1 + p_3 + p_5 + p_7 = \tfrac{1}{8}$

$P(A'B' = 1\,{-}1):\quad p_2 + p_6 + p_{10} + p_{14} = \tfrac{1}{8}$

The sum of the last three rows equals on one side to $\tfrac{3}{8}$

but also to $p_5 + p_6 + p_7 + p_8 +$ positive terms hence,

by the first row it is $\geq \tfrac{1}{2}$. Contradiction!

According to the existence theorem of SectionII each contextual EM has an equivalent HVM that is weakly but not strongly deterministic. To demonstrate this theorem, we list a weakly deterministic HVM which is equivalent to the EPR EM, as follows:

| $P(\lambda) = \tfrac{1}{8}$ | AB | A'B | AB' | A'B' |
|---|---|---|---|---|
| $\lambda = 0$ | 1,−1 | 1,1 | 1,1 | 1,1 |
| $\lambda = 1$ | 1,−1 | 1,−1 | 1,−1 | 1,1 |
| $\lambda = 2$ | 1,−1 | 1,−1 | 1,−1 | 1,1 |
| $\lambda = 3$ | 1,−1 | 1,−1 | 1,−1 | 1,−1 |
| $\lambda = 4$ | −1,1 | −1,1 | −1,1 | −1,1 |
| $\lambda = 5$ | −1,1 | −1,1 | −1,1 | −1,−1 |
| $\lambda = 6$ | −1,1 | −1,1 | −1,1 | −1,−1 |
| $\lambda = 7$ | −1,1 | −1,−1 | −1,−1 | −1,−1 |

To see that this HVM is not strongly deterministic note, for example, that for $\lambda = 1, 2$ the value of $B'$ is not determined and for $\lambda = 0, 7$ the value of $B$ is not determined etc.

This HVM is evidently weakly deterministic and hence is outcome independent, therefore it cannot be parameter independent otherwise it would be strongly deterministic.

It follows that this HVM model allows signaling. To see this fact assume that Alice can perform the measurements $A$ or $A'$ in a certain location and Bob can perform the measurements $B$ or $B'$ in another, space-like separated, location. Then if the system can be prepared in the state $\lambda = 1$, Alice can transmit the bit $1/-1$ to Bob who measures $B'$ just by choosing to measure $A/A'$ respectively.

**ACKNOWLEDGMENT**

This research was supported by the *Israel Science Foundation*, grant number 690/2021.